\documentclass[a4paper,12pt]{article}
\usepackage{amssymb,amsmath}
\usepackage{epsfig,fancyhdr}
\usepackage{a4}
\usepackage{parskip}
\usepackage{color,soul}
\usepackage{amsthm,amsxtra,amscd,relsize,multirow}
\usepackage[all,2cell]{xy}\UseAllTwocells

\newcommand{\sect}[1]{\setcounter{equation}{0}\section{#1}}

\def\N{{\mathcal N}}
\def\L{{\mathcal L}}

\def\L{{\mathcal L}}

\def\b{\beta}

\def\f{\phi}
\def\vf{\varphi}

\def\l{\lambda}

\def\o{\omega}

\def\cosh{\mathrm{cosh}}

\def\p{\partial}
\def\rb{\right}
\def\lb{\left}

\newcommand{\eq}[1]{\begin{equation} #1 \end{equation}}
\newcommand{\al}[1]{\begin{align} #1 \end{align}}
\newcommand{\ml}[1]{\begin{multline} #1 \end{multline}}

\def\cp{\mathbb {CP}^3}

\begin{titlepage}
\title{A note on strings in deformed $AdS_4\times \cp$: giant magnon and single spike solutions}

\author{
M. Schimpf and R.C. Rashkov\thanks{e-mail:
rash@hep.itp.tuwien.ac.at; 
also Dept of Physics, Sofia University, Bulgaria.}
\ \\ \ \\
${}^{\dagger}$ Institute for Theoretical Physics, \\ Vienna
University of Technology,\\
Wiedner Hauptstr. 8-10, 1040 Vienna, Austria 
}
\date{}
 \end{titlepage}

\begin{document}

\maketitle
\thispagestyle{fancy}

 \begin{abstract}
  In this paper we  study the solitonic string solutions of magnon and single spike type
in the beta-deformed $AdS_4\times\cp$ background. We find the dispersion relations which  are supposed to give the anomalous dimension of the gauge theory operators. 

 \end{abstract}

\sect{Introduction}

The string/gauge theory duality has attracted the attention of the high energy community for many
 years. As an example of such duality
AdS/CFT correspondence remains in the focus of the recent studies. A new exciting example of this duality has been added, namely duality between string in $AdS_4\times \cp$ and $\N=6$ superconformal Chern-Simons theory suggested by \cite{ABJM}. 
The ABJM model \cite{ABJM} is originally defined in M-theory and is believed to be holographically dual to M-theory on $AdS_4\times S^7/\mathbb Z_k$. Its reduction to string theory is believed to be dual 
to two Chern-Simons theories of level $k$
and $-k$, respectively, and each with gauge group $SU(N)$. The two pairs of
chiral superfields transform in the bi-fundamental representation
of $SU(N) \times SU(N)$ and the R-symmetry is $SU(4)$ as it should be for
 $\mathcal N=6$ supersymmetry.

The semi-classical string has played an important role in studying various aspects of the
$AdS_4/CFT_3$ correspondence \cite{Arutyunov:2008if}-\cite{Abbott:2008qd}\footnote{
Certainly we are not able to cite the huge list of papers contributing to the subject, we 
just quote the necessary minimum.}. 
The developments and successes in this
particular case suggest the methods and tools that should be used to investigate the new emerging duality. 
An important role in these studies plays the integrability. Superstrings on $AdS_4\times\cp$ as a coset were first
studied in \cite{Arutyunov:2008if} which opens the door for investigation of 
 integrable structures in the theory. 
Various properties on the gauge theory side and tests on the string theory side,
like rigid rotating strings, pp-wave limit, relations to spin chains,
as well as a pure spinor formulation,  have been considered 
\cite{Bagger:2006sk}-\cite{D'Auria:2008cw}. 
The complete type-IIA Green-Schwarz string action in  $AdS_4\times\cp$ superspace 
has been constructed in \cite{Gomis:2008jt}.

The theories with reduced supersymmetry are important not only conceptually but also for describing realistic physics. A very elegant technique for generating  solutions with reduced supersymmetry was proposed by \cite{Lunin:2005jy}. The method consists in T-duality along an isometry direction, followed by shift with a parameter $\b$ defining the deformation, and another T-duality called TsT- transformation. The method was further studied and developed in \cite{Frolov:2005dj}. On should note that the deformed theory has a richer structure of the vacua than its undeformed cousin. It is believed that the AdS/CFT correspondence persists after the deformation, thus providing important information.
In this paper we are studying the solitonic string solutions of magnon and single spike type
in the beta-deformed $AdS_4\times\cp$ background. The dispersion relations are supposed to give the anomalous dimension of the gauge theory operators. 

The paper is organized as follows. In the Introduction we give the basic concept of magnon and single spike string solutions and their dispersion relations. In the second section we present the corresponding solutions in the deformed $AdS_4\times\cp$ background and their dispersion relations. The summary of our results is presented in Conclusions.

\sect{About giant magnons and single spikes in $AdS_4\times \cp$ in short}

In this section we will give very brief review of the giant magnon and single spike string 
solutions in $AdS_4\times \cp$. The purpose is to set up the notations and write the solutions 
in the undeformed case for comparison to our results.

Let us start with the $AdS_4\times \cp$ metric and field content are given by
\al{
& ds^2_{IIA}=\frac{R^3}{k}\lb(\frac{1}{4}
ds^2_{AdS_4}+ds^2_{\cp}\rb) \notag \\
& ds^2_{\cp}=d\xi^2+\cos^2\xi\sin^2\xi\lb(d\psi+\frac{1}{2}\cos\theta_1d\vf_1-
\frac{1}{2}\cos\theta_2d\vf_2\rb)^2 \notag \\
& \qquad +\frac{1}{4}\cos^2\xi(d\theta_1^2+\sin^2\theta_1 d\vf_1^2)
 +\frac{1}{4}\sin^2\xi(d\theta_2^2+\sin^2\theta_2 d\vf_2^2),
& e^{2\Phi}=\frac{R^3}{k^3} \label{metric-fields} \\
& C_1=\frac{k}{2}\lb((\cos^2\xi-\sin^2\xi)d\psi+\cos^2\xi\cos\theta_1d\vf_1+\sin^2\xi\cos\theta_2
d\vf_2\rb), \notag \\
& F_2=k\lb(-\cos\xi\sin\xi\,d\xi\wedge(2d\psi+\cos\theta_1d\vf_1-\cos\theta_2d\vf_2)
\right. \notag \\
&\left.\qquad -\frac{1}{2}\cos^2\xi\sin\theta_1d\theta_1\wedge d\vf_1-\frac{1}{2}\sin^2\xi
\sin\theta_2 d\theta_2\wedge d\vf_2 \rb) \notag \\
& F_4=-\frac{3R^3}{8}\o_{AdS_4}.
}

The general solitonic string solution in this background is very difficult to obtain, however it can be done in a certain subsectors. Let us describe below one of them.

Let us consider the case of $\theta_1=\theta_2=\pi/2$ and $\psi=0$ (see for instance \cite{Ryang:2008rc}). One can check that this is
a consistent truncation and for this subsector the Polyakov Lagrangian 
has the form 
\eq{
\L\sim -\frac{1}{4}\dot t^2 +\dot\xi^2-\xi'^2+\frac{1}{4}\cos^2\xi\:\big(
\dot\f_1^2-{\f'_1}^2\big) +\frac{1}{4}\sin^2\xi\:\big(
\dot\f_2^2-{\f'_2}^2\big)
\label{larg-undef-1}
}
The ansatz for a rotating string soliton is
\eq{
t=\kappa\tau, \quad \xi=\xi(y), \quad \f_1=\o_1+f_1(y),
\quad \f_1=\o_2+f_2(y), \quad y=c\sigma -d\tau.
}
The Virasoro constraints plus requirement of existing of a turning point at $\xi=\pi/2$ lead to the equation of motion for $\xi$
\eq{
\xi'^2=\frac{1}{4(c^2-d^2)}\big[(c^2)d^2)\kappa^2-\frac{C^2}{\sin^2\xi}
-c^2(\o_1^2\cos^2\xi+\o_2^2\sin^2\xi)  \big],
\label{undef-eom-1}
}
where $\o_2 C=\kappa^2 d$. The turning points can be obtained from \eqref{undef-eom-1}
with one of them $\xi_-=\pi/2$ and the other,$\xi_+$, determined by
\eq{
(\kappa^2-\o_2^2\big(c^2-\frac{\kappa^2 d^2}{\o_2^2}\big)=0.
}
The two roots correspond to magnon and single spike solutions as follows
\eq{
\sin^2\xi_+=\frac{d^2}{(\o_2^2-\o_1^2)c^2} \,\,\text{(giant magnon)}, \quad
\sin^2\xi_+=\frac{c^2}{(\o_2^2-\o_1^2)d^2} \,\,\text{(spike)}
\label{turning-g-sp}
}
The dispersion relation in the case of $R_t\times \cp$ is obtained to be\cite{Ryang:2008rc}
\eq{
E-J_1=\sqrt{J_2^2+\frac{8\l}{\sin^2\frac{p}{4}}},
\label{disp-giant-1}
}
which is half of the standard one obtained from Neumann-Rosochatius integrable system
in $R_t\times S^3\times S^3$ \cite{Ahn:2008hj}. Here the magnon momentum is identified with the angle difference $\Delta\f_2$.

In the case of single spike string solution the dispersion relation is obtained in the form
\eq{
J_2=\sqrt{J_1^2+8\l\sin^2\bar\xi},
}
where $\bar\xi=\pi/2-\xi_+$. The large energy is combined with the large winding number around
$\Delta\f_2$ and they are related as follows
\eq{
E-\frac{\sqrt{2\l}}{2}\Delta\f_2=2\sqrt{2}\big(\frac{\pi}{2}-\xi_+\big)
=2\sqrt{2}\:\bar\xi.
}

With these results in the subspace of $\theta_1=\theta_2=\pi/2$ and $\psi=0$ we conclude the short review of the solitonic solutions in $AdS_4\times \cp$. In the next section we consider the deformed case.

\sect{Giant magnons and single spikes in deformed $AdS_4\times \cp$}

Let us start with the deformed $AdS_4\times \cp$ metric and the field content 
\cite{Imeroni:2008cr}
\al{
& ds^2_{IIA}=\frac{R^3}{k}\lb(\frac{1}{4}
ds^2_{AdS_4}+ds^2_{\widetilde{\cp}}\rb),\quad e^{2\Phi}=\frac{R^3}{k^3}G  \label{metric-fields1} \\
& ds^2_{\widetilde{\cp}}=d\xi^2
+\frac{1}{4}\sin^2\xi(d\theta_2^2+G\sin^2\theta_2 d\vf_2^2)
+\frac{1}{4}\cos^2\xi(d\theta_1^2+G\sin^2\theta_1 d\vf_1^2) \notag \\
& +G\cos^2\xi\sin^2\xi\lb(d\psi+\frac{1}{2}\cos\theta_1d\vf_1-
\frac{1}{2}\cos\theta_2d\vf_2\rb)^2 
+\tilde\gamma^2 G\cos^4\xi\sin^4\xi\sin^2\theta_1\sin^2\theta_2\, d\psi^2, 
\notag
}
\al{ 
& B=-\frac{\tilde\gamma G R^3}{k}\cos^2\xi\sin^2\xi \,\,\times \notag \\
& \qquad \times\lb(\frac{1}{2}\cos^2\xi\sin^2\theta_1\cos\theta_2\,d\psi\wedge d\vf_1+
\frac{1}{2}\sin^2\xi\sin^2\theta_2\cos\theta_1\,d\psi\wedge d\vf_2 \right. \notag \\
& \left.\qquad +\frac{1}{4}(\sin^2\theta_1\sin^2\theta_2+\cos^2\xi\sin^2\theta_1\cos^2\theta_2
+\sin^2\xi\sin^2\theta_2\cos^2\theta_1)\,d\vf_1\wedge d\vf_2\rb) \notag \\
& F_2=k\lb(-\cos\xi\sin\xi\,d\xi\wedge(2d\psi+\cos\theta_1d\vf_1-\cos\theta_2d\vf_2)
\right. \notag \\
&\left.\qquad -\frac{1}{2}\cos^2\xi\sin\theta_1\,d\theta_1\wedge d\vf_1-\frac{1}{2}\sin^2\xi
\sin\theta_2\, d\theta_2\wedge d\vf_2 \rb) \notag \\
& F_4=-\frac{3R^3}{8}(\omega_{AdS_4}+4\tilde\gamma\cos^3\xi\sin^3\xi\sin\theta_1\sin\theta_2\,
d\xi\wedge d\psi\wedge d\theta_1 \wedge d\theta_2) \notag \\
& \qquad -\frac{R^3}{8}d(\tilde\gamma G\cos^2\xi\sin^2\xi
(\cos^2\xi\sin^2\theta_1-\sin^2\xi\sin^2\theta_2))\wedge d\psi\wedge d\vf_1\wedge d\vf_2,
\label{def-fields}
}
where $\tilde\gamma = (R^3/4k)\gamma$ and
\eq{
G^{-1}=1+\tilde\gamma^2\cos^2\xi\sin^2\xi(\sin^2\theta_1\sin^2\theta_2+
\cos^2\xi\sin^2\theta_1\cos^2\theta_2+\sin^2\xi\sin^2\theta_2\cos^2\theta_1).
}
One can see that the equations  \eqref{metric-fields1} and \eqref{def-fields} preserve four supercharges and thus matching  the dual $\N=2$ supersymmetric gauge theory in three dimensions.
According to \cite{Leigh:1995ep} the gauge theory superpotenital is modified and takes the form
\eq{
\frac{4\pi}{k}\mathrm{Tr}\big(A_1B_1A_2B_2-A_1B_2A_2B_1\big)\:
\rightarrow\:
\frac{4\pi}{k}\mathrm{Tr}\big(e^{-i\pi\gamma/2}A_1B_1A_2B_2-e^{i\pi\gamma/2}A_1B_2A_2B_1\big).
\notag
}

To find solitonic string solutions we employ the followinf ansatz
\al{
& t=\kappa\tau; \quad \theta_1=\theta_2=\theta=\frac{\pi}{2}; \quad \psi=0\notag \\
& \xi=\xi(y); \quad \vf_i=\o_i\tau +f_i(y)
}
where $y=c\sigma-d\tau$.

\subsection{Equations and solutions in the deformed theory}

Let us write first the Lagrangian setting $\theta_1=\theta_2=\theta$
\ml{
\L\sim \frac{1}{4}\dot t^2-\dot\xi^2+{\xi'}^2+
G\cos^2\xi\sin^2\xi(1+\tilde\gamma^2\cos^2\xi\sin^2\xi\sin^4\theta)(-\dot\psi^2+{\psi'}^2)\\
+\frac{G}{4}\cos^2\xi(\sin^2\theta+\sin^2\xi\cos^2\theta)(-\dot\vf_1^2+{\vf_1'}^2)
+\frac{G}{4}\sin^2\xi(\sin^2\theta+\cos^2\xi\cos^2\theta)(-\dot\vf_2^2+{\vf_2'}^2) \\
+\frac{1}{4}(-\dot\theta^2+{\theta'}^2)+G\cos^2\xi\sin^2\xi\cos\theta(-\dot\psi(\dot\vf_1-\dot\vf_2)
+{\psi'}(\vf_2'-\vf_2'))\\
-\frac{G}{2}\cos^2\xi\sin^2\xi\cos^2\theta(-\dot\vf_1\dot\vf_2+\vf_1'\vf_2')\\
-2\tilde\gamma G\cos^2\xi\sin^2\xi\lb[\cos^2\xi\sin^2\theta\cos\theta(-\dot\psi(\dot\vf_1+\dot\vf_2)+
\psi'(\vf_1'+\vf_2'))\right. \\
\left. +\frac{\sin^2\theta}{4}(\sin^2\theta+2\cos^2\xi\cos^2\theta)(-\dot\vf_1\vf'_2
+\vf_1'\dot\vf_2)\right].
\label{lagrangian-def-1}
}
Examining the equations of motion for $\theta$ and $\psi$ one can see that 
$\theta=\frac{\pi}{2}$ and $\psi=0$ are solutions to the equations of motion.

Then  we proceed with the choice of $\theta=\frac{\pi}{2}; \quad \psi=0$.
One can see that with this ansatz the B-field becomes (we set for a moment $R^3/\kappa=1$)
\eq{
B=-\frac{\tilde\gamma G}{4}\cos^2\xi\sin^2\xi\,d\vf_1\wedge d\vf_2.
\label{B-field1}
}
The metric takes the form
\eq{
ds^2_{\widetilde{\cp}}(\theta=\frac{\pi}{2},\psi=0)
=d\xi^2+\frac{G}{4}\cos^2\xi\,d\vf_1^2+\frac{G}{4}\sin^2\xi\,d\vf_2^2,
}
where
\eq{
G^{-1}=1+\tilde\gamma^2\cos^2\xi\sin^2\xi.
}

We make further assumptions for the fields making a rotating string solitonic ansatz
\eq{
t=\kappa\tau, \quad \vf_1=\omega_1\tau+ f_1(y), \quad \vf_2=\omega_2\tau +f_2(y),
\quad y=c\sigma-d\tau.
}
The Lagrangian we will work with takes the form 
\ml{
\L\sim \frac{1}{4}\dot t^2+\big(c^2-d^2\big){\xi'}^2
+\frac{G}{4}\cos^2\xi\Big(-\big(\o_1-df'_1\big)^2+c^2{f'_1}^2\Big)\\
+\frac{G}{4}\sin^2\xi\Big(-\big(\o_2-df'_2\big)^2+c^2{f'_2}^2\Big)
-\frac{G}{2}\tilde\gamma\cos^2\xi\sin^2\xi
\Big(-\big(\o_1-df'_1\big)cf_2'+\big(\o_2-df'_2\big)cf_1'\Big)
\label{act-redI1}
}

\paragraph{Virasoro constraints:}

 Next step is to elaborate the conditions imposed by
the two Virasoro constraints:

\textbf{a)} $T_{\tau\tau}+T_{\sigma\sigma}=0$
\eq{
(c^2+d^2)\xi'^2+\frac{G}{4}\cos^2\xi\Big[(\o_1-df'_1)^2+c^2{f'_1}^2\Big]
+\frac{G}{4}\sin^2\xi\Big[(\o_2-df'_2)^2+c^2{f'_2}^2\Big]=\frac{\kappa^2}{4}
\label{vir-diag0}
}

\textbf{b)} $T_{\tau\sigma}=0$
\eq{
-cd\xi'^2+\frac{G}{4}\cos^2\xi(\o_1-df'_1)cf'_1+\frac{G}{4}\sin^2\xi (\o_2-df'_2)cf'_2=0.
}
or
\eq{
\xi'^2+\frac{G}{4}\cos^2\xi\Big[{f'_1}^2+\frac{\o_1^2-2\o_1df'_1}{c^2+d^2}\Big]
+\frac{G}{4}\sin^2\xi\Big[{f'_2}^2+\frac{\o_2^2-2\o_2df'_2}{c^2+d^2}\Big]=
\frac{\kappa^2}{4(c^2+d^2)}.
\label{vir-diag}
}
and 
\eq{
-\xi'^2+\frac{G}{4}\cos^2\xi\Big[\frac{\o_1f'_1}{d}-{f'_1}^2\Big]
+\frac{G}{4}\sin^2\xi\Big[\frac{\o_1f'_1}{d}-{f'_1}^2\Big]=0.
\label{vir-off}
}
Adding the last two equations we find
\ml{
\frac{G}{4}\o_1\cos^2\xi\Big[\frac{\o_1}{c^2+d^2}+\frac{(c^2-d^2)f'_1}{d(c^2+d^2)}\Big]
+\frac{G}{4}\o_2\sin^2\xi\Big[\frac{\o_2}{c^2+d^2}\\
+\frac{(c^2-d^2)f'_2}{d(c^2+d^2)}\Big]
=\frac{\kappa^2}{4(c^2+d^2)}.
\label{vir-comb}
}

\paragraph{Equations of motion:}

In contrast to the undeformed case here one must account for the non-vanishing B-field (see the Lagrangian \eqref{act-redI1}). Then we find
\eq{
\p_y\Big\{\frac{G}{2}\cos^2\xi\Big[(\o_1-df'_1)d+c^2f'_1\Big]
-\frac{G}{2}\tilde\gamma\cos^2\xi\sin^2\xi\Big[(\o_2-df'_2)c+cd{f'_2}^2\Big]\Big\}=0
}
or
\eq{
\p_y\Big\{\frac{G}{2}\sin^2\xi\Big[(\o_1-df'_1)d+c^2f'_1\Big]
-\frac{G}{2}\tilde\gamma\cos^2\xi\sin^2\xi\,\o_2 c\Big\}=0.
}
Analogously
\eq{
\p_y\Big\{\frac{G}{2}\sin^2\xi\Big[(\o_2-df'_2)d+c^2f'_2\Big]
+\frac{G}{2}\tilde\gamma\cos^2\xi\sin^2\xi\,\o_1 c\Big\}=0.
}
From here we find
\al{
& (c^2-d^2)f'_1=\frac{2A_1}{G\cos^2\xi}-\o_1 d+\tilde\gamma\o_2c\,\sin^2\xi,
\label{eom1}\\
& (c^2-d^2)f'_2=\frac{2A_2}{G\sin^2\xi}-\o_2 d-\tilde\gamma\o_1c\,\cos^2\xi.
\label{eom2}
}
Substituting this expression into \eqref{vir-comb} we find 
\eq{
2A_1\o_1+2A_2\o_2=\kappa^2d. 
\label{rel-const}
}

The substitution into the first Virasoro constraint gives the equation for $\xi$
\ml{
\xi'^2=\frac{1}{(c^2-d^2)^2}\Big\{\big(c^2+d^2)\kappa^2-\frac{4A_1^2}{G\cos^2\xi}
-\frac{4A_2^2}{G\sin^2\xi}-c^2\o_1^2\cos^2\xi-c^2\o_2^2\sin^2\xi  \\
 +4c\tilde\gamma\big(A_2\o_1\cos^2\xi-A_1\o_2\sin^2\xi\big)\Big\}
\label{eq-fin0-1case}
}


The solutions we are looking for are of folded type and therefore one is 
to require a turning point at $\xi=\frac{\pi}{2}$. This condition forces
\eq{
\xi=\frac{\pi}{2}\quad \Rightarrow \quad A_1=0.
}
In order $\xi=\frac{\pi}{2}$ to be a turning point it must be zero of 
\eqref{eq-fin0-1case}. This gives (we use also that $A_1=0$ and that 
$2A_2\o_2=\kappa^2\o_2$)
\eq{
\Big(\frac{\kappa^2}{\o_2^2}-1\Big)\Big(\frac{\kappa^2}{\o_2^2}-
\frac{c^2}{d^2}\Big)=0
\quad \Rightarrow \quad
\begin{cases}
 \kappa^2=\o_2^2\quad \text{magnon}\\
d^2\kappa^2=c^2\o_2^2 \quad \text{spike}
\end{cases}
}
Therefore we have two cases: magnon solutions and single spike solutions.

The solutions for $f'_1$ and $f'_2$ become
\al{
& (c^2-d^2)f'_1=-\o_1 d+\tilde\gamma\o_2c\,\sin^2\xi,
\label{eom1-2}\\
& (c^2-d^2)f'_2=\frac{2A_2}{G\sin^2\xi}-\o_2 d-\tilde\gamma\o_1c\,\cos^2\xi.
\label{eom2-2}
}

We are interested in the dispersion relation for the obtained solutions. As it is well known from the $AdS_5\times S^2$ background, we have in general two types interesting soliton solutions - giant magnons and spiky strings. The shape of the solutions depends on the parameters and here we will consider the two cases separately.

\paragraph{Giant magnon solutons:}

Since we already fixed $A_1$ to be vanishing, from \eqref{rel-const} we find
\eq{
A_2=\frac{d\kappa^2}{2\o_2}
}
The magnon type solutions are determined by
\eq{
\kappa^2=\o_2^2.
}
The equation for $\xi$ takes the form
\ml{
\xi'^2  =\frac{1}{4(c^2-d^2)^2}\Big\{\big(c^2+d^2)\kappa^2
-\frac{4A_2^2}{G\sin^2\xi}-c^2\big(\o_1^2\cos^2\xi+\o_2\sin^2\xi\big)  \\
 +4c\tilde\gamma A_2\o_1\cos^2\xi\Big\}
}
We substitute $\kappa_2=\o_2^2$ and $A_2=d\kappa^2/(2\o_2)$ and obtain
\eq{
\xi'^2=\frac{c^2\Omega_0^2\cos^2\xi}{4(c^2-d^2)^2\sin^2\xi}\Big\{\sin^2\xi
-\sin^2\xi_0\Big\},
\label{eq-mag-1}
}
where we used the following notations
\eq{
 \Omega_0^2=\Big(\o_2^2-\big(\o_1-\frac{d\tilde\gamma}{c}\o_2\big)^2\Big), \quad
 \sin\xi_0^2=\frac{d^2\o_2^2}{c^2\Omega_0^2}.
}

The solution of the equation \eqref{eq-mag-1} is
\eq{
\cos\xi=\frac{\cos\xi_0}{\cosh\big(\frac{c\Omega_o\cos\xi_0}{4(c^2-d^2)}y\big)}.
\label{eq-mag-sol1}
}

\paragraph{Single spike solutions:}

The single spike type solutions are determined by
\eq{
d^2\kappa^2=c^2\o_2^2.
}

The equation of motion for $\xi$ becomes
\eq{
\xi'^2=\frac{c^2\Omega_0^2\cos^2\xi}{4(c^2-d^2)^2\sin^2\xi}\Big\{\sin^2\xi
-\sin^2\xi_0\Big\},
\label{eq-spike-1}
}
where we used the following definitions
\eq{
 \Omega_0^2=\Big(\o_2^2-\big(\o_1-\frac{c\tilde\gamma}{d}\o_2\big)^2\Big), \quad
 \sin\xi_0^2=\frac{c^2\o_2^2}{d^2\Omega_0^2}.
}


\subsection{The dispersion relations}

Let us start with the conserved charges. In the case of giant magnons 
the momenta are given by
\eq{
P_{\f_1}=\frac{\cos^2\xi}{2(1-v^2)}\big[\o_1-\tilde\gamma v \o_2\big],
\quad
P_{\f_2}=\frac{\kappa}{2}-\frac{\o_2\cos^2\xi}{2(1-v^2)}\big[\o_1-\tilde\gamma v \o_2\big],
\quad P_t=\frac{\kappa}{2}.
\label{j2-mag1}
}

For the sting configuration of single spike type we have large winding number which is combined with the energy to produce finite result.
The momentum along $\f_1$ is the very same
\eq{
P_{\f_1}=\frac{\cos^2\xi}{2(1-v^2)}\big[\o_1-\tilde\gamma v \o_2\big],
\label{j1-sp1}
}
while the second momentum is
\eq{
P_{\f_2}=\frac{\o_2\cos^2\xi}{2(1-v^2)}.
\label{j2-sp1}
}
Note that both momenta are finite.

Once we have the expressions for the conserved quantities, it is easy to obtain the corresponding dispersion relations.

\paragraph{Giant magnon case:}

The finite combination is
\eq{
P_t-P_{\f_2}=\frac{\o_2\cos^2\xi}{2(1-v^2)}\big[\o_1-\tilde\gamma v \o_2\big].
\label{disp-mag1}
}

Integrating the above expression we obtain
\eq{
E-J_2=\frac{\sqrt{\l}}{\pi}\frac{\o_2\cos\xi_0}{\Omega_0}.
\label{disp-mag2}
}
The finite spin $J_1$ is
\eq{
J_1=\frac{\sqrt{\l}}{\pi}\frac{\cos\xi_0}{\Omega_0}\big(\o_1-\tilde\gamma v\o_2\big).
\label{disp-mag3}
}
From here we easily obtain the dispersion relation
\eq{
E-J_2=\sqrt{J_1^2-\frac{\l}{\pi^2}\sin^2\big(\frac{p}{2}-\pi\b\big)},
\label{disp-mag4}
}
where we made the identification $\cos\xi_0=\sin(p/2-\pi\b)$.

The angle deficit is
\eq{
\frac{\Delta\vf_2}{2}=\big(\frac{\pi}{2}-\xi_0\big)
-\tilde\gamma\frac{\sqrt{\o_2^2-\Omega_0^2}}{\Omega_0}
\cos\xi_0.
\label{disp-mag5}
}

\paragraph{Single spike case:}

The momenta $J_1$ and $J_2$ are finite. Integrating \eqref{j1-sp1} and
\eqref{j2-sp1} and using the relations between them we find
\eq{
J_2=\sqrt{J_1^2-\frac{\l}{\pi^2}\sin^2\big(\frac{p}{2}-\pi\b\big)},
}
where we made the identification $\cos\xi_1=\sin(p/2-\pi\b)$. 

Both, the energy and the angle deficit are divergent, but the combination
$E-\Delta\vf_2$ is finite and is given by
\eq{
E-T\Delta\vf_2=\frac{\sqrt{\l}}{\pi}\big(\frac{\pi}{2}-\xi_1\big)
-\tilde\gamma\frac{\sqrt{\l}}{\pi}\frac{\sqrt{\o_2^2-\Omega_1^2}}{\Omega_1}\cos\xi_1.
}

We conclude by  stressing on the analogy with the undeformed case, which can be recovered by taking $\tilde\gamma \rightarrow 0$.

\sect{Conclusions}

In this short note we studied the existence of giant magnon and single spike string solutions in beta-deformed $AdS_4\times \cp$ background. 
We used the representation of the background an $U(1)$ fibration over $S^2\times S^2$ with a fiber coordinate $\psi$ and the two spheres described by $\f_i, \theta_i, i=1,2$. The deformation 
via solution generating TsT (T-duality, shift, T-duality) technique is known and is usually used to construct theories with less supersymetry. In this paper we find giant magnon and single spike classical string solutions in the deformed background We find their dispersion relations which are supposed to describe the anomalous dimensions of certain class gauge theory operators. The complete map between the results obtained from string side and gauge theory deserves further careful study.

\bigskip
\leftline{\bf Acknowledgments}
\smallskip
We thank H.Dimov and Max Kreuzer for useful discussion at an early stage of the project.
R.R. acknowledges the warm hospitality and stimulating atmosphere at the ITP of TU Wien.
This work was supported in part by the Austrian Research Fund FWF grant \# I192.
R.R. is partly supported by the BNSF grant VU-F-201/06 and DO 02-257.



\begin{thebibliography}{99}


\bibitem{Bagger:2006sk}
J.~Bagger and N.~Lambert, {{\em Phys. Rev.} {\bf  D75} (2007)
045020}, {{ arXiv:hep-th/0611108}}.
A.~Gustavsson, {arXiv:0709.1260 [hep-th]}.
J.~Bagger and N.~Lambert, {{\em Phys. Rev.} {\bf  D77} (2008)
065008}, {{\tt arXiv:0711.0955 [hep-th]}}.
J.~Bagger and N.~Lambert,{{\em JHEP} {\bf 02}  (2008)  105}, {{\tt
arXiv:0712.3738 [hep-th]}}.

\bibitem{Schwarz:2004yj}J.~H. Schwarz,{{\em  JHEP} {\bf 11} (2004)  078}, 
{{arXiv:hep-th/0411077}}.

\bibitem{ABJM}
O. Aharony, O. Bergman, D.~L. Jafferis and J. Maldacena, 
JHEP {\bf 0810} (2008) 091 [arXiv:hep-th/0806.1218].

\bibitem{Hofman:2006xt}
  D.~M.~Hofman and J.~M.~Maldacena,
  J.\ Phys.\ A  {\bf 39} (2006) 13095
  [arXiv:hep-th/0604135].

\bibitem{Kruczenski:2006pk}
  M.~Kruczenski, J.~Russo and A.~A.~Tseytlin,
  JHEP {\bf 0610} (2006) 002
  [arXiv:hep-th/0607044].




\bibitem{Arutyunov:2008if}
  G.~Arutyunov and S.~Frolov,
  arXiv:0806.4940 [hep-th].

\bibitem{Stefanski:2008ik}
  B.~.~j.~Stefanski,
  Nucl.\ Phys.\  B {\bf 808} (2009) 80
  [arXiv:0806.4948 [hep-th]].

\bibitem{McLoughlin:2008ms}
  T.~McLoughlin and R.~Roiban,
  arXiv:0807.3965 [hep-th].

\bibitem{Alday:2008ut}
  L.~F.~Alday, G.~Arutyunov and D.~Bykov,
  JHEP {\bf 0811} (2008) 089
  [arXiv:0807.4400 [hep-th]].
\bibitem{Krishnan:2008zs}
  C.~Krishnan,
  JHEP {\bf 0809} (2008) 092
  [arXiv:0807.4561 [hep-th]].




\bibitem{Benna:2008zy}
  M.~Benna, I.~Klebanov, T.~Klose and M.~Smedback,
  arXiv:0806.1519 [hep-th].


\bibitem{Grignani:2008is}
  G.~Grignani, T.~Harmark and M.~Orselli,
  arXiv:0806.4959 [hep-th].

\bibitem{Armoni:2008kr}
  A.~Armoni and A.~Naqvi,
  arXiv:0806.4068 [hep-th].

\bibitem{Chen:2008qq}
  B.~Chen and J.~B.~Wu,
  arXiv:0807.0802 [hep-th].

\bibitem{Bonelli:2008us}
  G.~Bonelli, P.~A.~Grassi and H.~Safaai,
  arXiv:0808.1051 [hep-th].


\bibitem{D'Auria:2008cw}
  R.~D'Auria, P.~Fre', P.~A.~Grassi and M.~Trigiante,
  arXiv:0808.1282 [hep-th].

\bibitem{Gomis:2008jt}
  J.~Gomis, D.~Sorokin and L.~Wulff,
  JHEP {\bf 0903} (2009) 015
  [arXiv:0811.1566 [hep-th]].


\bibitem{Rashkov:2008rm}
  R.~C.~Rashkov,
  Phys.\ Rev.\  D {\bf 78} (2008) 106012
  [arXiv:0808.3057 [hep-th]].

\bibitem{Ahn:2008hj}
  C.~Ahn, P.~Bozhilov and R.~C.~Rashkov,
  JHEP {\bf 0809} (2008) 017
  [arXiv:0807.3134 [hep-th]].

 
\bibitem{Ryang:2008rc}
  S.~Ryang,
  JHEP {\bf 0811} (2008) 084
  [arXiv:0809.5106 [hep-th]].


\bibitem{Kalousios:2009mp}
  C.~Kalousios, M.~Spradlin and A.~Volovich,
  JHEP {\bf 0907} (2009) 006
  [arXiv:0902.3179 [hep-th]].

\bibitem{Kalousios:2009ey}
  C.~Kalousios, C.~Vergu and A.~Volovich,
  arXiv:0905.4702 [hep-th].

\bibitem{Papathanasiou:2009en}
  G.~Papathanasiou and M.~Spradlin,
  JHEP {\bf 0907} (2009) 036
  [arXiv:0903.2548 [hep-th]].

\bibitem{Abbott:2009um}
  M.~C.~Abbott, I.~Aniceto and O.~O.~Sax,
  arXiv:0903.3365 [hep-th].

\bibitem{Abbott:2008qd}
  M.~C.~Abbott and I.~Aniceto,
  arXiv:0811.2423 [hep-th].

\bibitem{Bombardelli:2008qd}
  D.~Bombardelli and D.~Fioravanti,
  JHEP {\bf 0907} (2009) 034
  [arXiv:0810.0704 [hep-th]].


\bibitem{Leigh:1995ep}
  R.~G.~Leigh and M.~J.~Strassler,
  Nucl.\ Phys.\  B {\bf 447} (1995) 95
  [arXiv:hep-th/9503121].

\bibitem{Lunin:2005jy}  O.~Lunin and J.~M.~Maldacena,
  JHEP {\bf 0505}, 033 (2005), [arXiv:hep-th/0502086].

\bibitem{Frolov:2005dj}
  S.~Frolov,
  JHEP {\bf 0505} (2005) 069
  [arXiv:hep-th/0503201].



\bibitem{Bobev:2007bm}
  N.~P.~Bobev and R.~C.~Rashkov,
  Phys.\ Rev.\  D {\bf 76}, 046008 (2007)
  [arXiv:0706.0442 [hep-th]].

\bibitem{Bobev:2006fg}
  N.~P.~Bobev and R.~C.~Rashkov,
  Phys.\ Rev.\  D {\bf 74}, 046011 (2006)
  [arXiv:hep-th/0607018].

\bibitem{Chu:2006ae}
  C.~S.~Chu, G.~Georgiou and V.~V.~Khoze,
  JHEP {\bf 0611} (2006) 093
  [arXiv:hep-th/0606220].


\bibitem{Bykov:2008bj}
  D.~V.~Bykov and S.~Frolov,
  JHEP {\bf 0807} (2008) 071
  [arXiv:0805.1070 [hep-th]].

\cite{Imeroni:2008cr}
\bibitem{Imeroni:2008cr}
  E.~Imeroni,
  JHEP {\bf 0810} (2008) 026
  [arXiv:0808.1271 [hep-th]];

  N.~Akerblom, C.~Saemann and M.~Wolf,
  arXiv:0906.1705 [hep-th].



\end{thebibliography}
\end{document}